\documentclass{article}
\usepackage{spconf,amsmath,graphicx}

\title{MAXIMIZING AUDIO EVENT DETECTION MODEL PERFORMANCE ON SMALL DATASETS THROUGH KNOWLEDGE TRANSFER, DATA AUGMENTATION, AND PRETRAINING: AN ABLATION STUDY}
\name{Daniel Tompkins*\thanks{Corresponding author: daniel.tompkins@microsoft.com}, Kshitiz Kumar, Jian Wu}
\address{Microsoft}
\begin{document}
%
\maketitle
\begin{abstract}
An Xception model reaches state-of-the-art (SOTA) accuracy on the ESC-50 dataset for audio event detection through knowledge transfer from ImageNet weights, pretraining on AudioSet, and an on-the-fly data augmentation pipeline. This paper presents an ablation study that analyzes which components contribute to the boost in performance and training time. A smaller Xception model is also presented which nears SOTA performance with almost a third of the parameters. 
\end{abstract}
\section{Introduction}
\label{sec:intro}

Audio Event Detection (AED) has greatly benefited from deep-learning methods with CNN-based models and, more recently, Transformer-based models providing significant increases in classification accuracy and raising the state-of-the-art (SOTA). However, many AED datasets have a small number of labeled examples, especially when compared to other domains such as vision and language. This limits the ability of large models to train directly on datasets without over- or under-fitting. 

Solutions to the small data problem include data augmentation, knowledge transfer, and pretraining on a larger labeled dataset such as AudioSet~\cite{audiosetontology} or a self-supervised learning approach with substantial amounts of unlabeled data~\cite{zhang2021bigssl}. However, further study is necessary to determine which of these techniques or technique combinations provide the optimal solution. This paper analyzes the effects of knowledge transfer, data augmentation, and pretraining on a popular small dataset, ESC-50~\cite{esc50}. We train two models, Xception and Xception-small, and conduct an ablation study of different combinations of the mentioned techniques and record the ESC-50 accuracy. 

Our Xception model with knowledge transfer from ImageNet weights, pretraining on AudioSet, with on-the-fly data augmentation reaches SOTA while our Xception-small model reaches near SOTA despite being much smaller, providing an option for low-compute scenarios.

\section{Previous Work}
\label{sec:previouswork}

The development of Pretrained Audio Neural Networks (PANNS) has presented impact of different architectures, augmentation, and other training and dataset options and how it impacts AudioSet tagging~\cite{kong2020panns}. The PANNS study also evaluated on ESC-50 and reached high quality. Other studies of data augmentation for AED and other audio-tagging tasks include \cite{wei2018sample, takahashi2016deep, xu2018mixupxception}. Some tools have become standard for augmenting data and creating synthetic data for AED, most notably \cite{salamon2017scaper}.

The Audio Spectrogram Transformer (AST) paper applies ImageNet weights to a ViT-type architecture for audio~\cite{gong2021ast}. Their evaluation on the ESC-50 dataset is currently SOTA, and they found applying ImageNet weights significantly improved their results. However, they implied this improvement might only be available for very large models. Initializing models with ImageNet-trained weights has also been explored in~\cite{palanisamy2020rethinking, guzhov2021esresnet, gwardys2014deep} for CNN-based models.

There have been several other models trained on large datasets, AudioSet or other large dataset, such as ~\cite{openl3, vggish}. Recently, a wav2vec approach with a large billion-parameter model was trained with self-supervised learning (SSL) on unlabeled data that has been used as a general embedding model for AED and several other speech tasks~\cite{zhang2021bigssl}. A near SOTA of ESC-50 was reached in \cite{kumar2020sequential} by using a sequential self-teaching model which provided a significant SOTA lead in the ESC-50 leaderboard at the time of its publication. 

Xception models have also been used for AED, such as in \cite{xu2018mixupxception, gajarsky2018xception}, as the model provides high performance with relatively few parameters compared to other top-performing models. 

Our paper builds on the work described above but focuses on an ablation study of how each component---data augmentation, knowledge transfer with ImageNet, pretraining with AudioSet---impacts performance, with a specific focus on small datasets (ESC-50) and relatively small models (Xception and Xception-small).

\section{Datasets}
\label{sec:pagestyle}

We focus our evaluation metrics on the ESC-50 dataset~\cite{esc50} because it has a comprehensive and current leaderboard and because it has relatively few examples per class. The ESC-50 dataset is structured in 5 folds, so we trained and evaluated our models 5 times, rotating the withheld fold for evaluation. We average the evaluation folds to obtain our accuracy scores. ESC-50 contains 50 balanced classes, 8 files of each class per fold for a total of 2000 files. Each example is exactly five seconds long and contains only one class. 

For pretraining, we use AudioSet's unbalanced dataset ~\cite{audiosetontology}, which contains over 2 million labeled examples covering 527 classes, which are structured as a hierarchical ontology. Most examples are 10 seconds in length and contain multiple labels. We use AudioSet's evaluation dataset to evaluate the model after each epoch of training on the unbalanced dataset.

\begin{figure}[htb]

  \centering
  \centerline{\includegraphics[width=9.0cm]{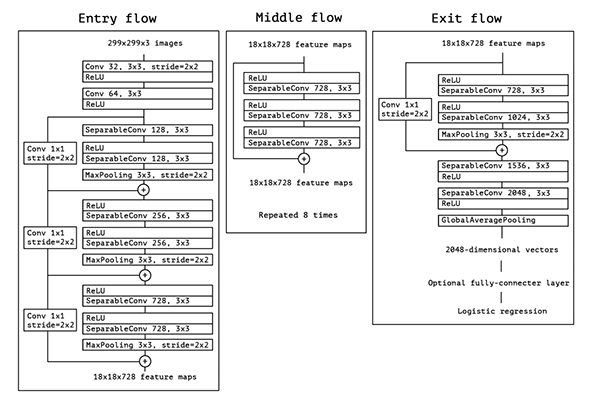}}


\caption{The Xception architecture, as described in \cite{chollet2017xception}. Our Xception-small model does not repeat the middle flow.}
\label{fig:xception}
\end{figure}

\section{Knowledge Transfer and Augmentation}
\label{sec:typestyle}

In the absence of a large labeled AED dataset, knowledge transfer from another domain and data augmentation are common ways of approaching the problem small datasets such as ESC-50 pose to deep learning models. 

\subsection{ImageNet Weight Initialization}
\label{ssec:imagenet}

The Xception model has been trained with ImageNet~\cite{krizhevsky2012imagenet}, which is a large dataset of millions of images. Several previous studies \cite{palanisamy2020rethinking, guzhov2021esresnet, gwardys2014deep, gong2021ast} have applied ImageNet weights to models as a replacement for weight initialization. Despite the difference between image data and audio data, using ImageNet weights is often shown to give a performance increase. To apply ImageNet weights to the Xception model, we used a method similar to \cite{gong2021ast}, which includes averaging the first three channels of the model input into one channel (RGB image channels to one spectrogram channel). 

We attempted to apply ImageNet weights to the Xception-small model by starting with the full Xception model with ImageNet weights and deleting the middle flow repetitions, but we did not see any boost in performance. Further study is needed on methods of altering a pretrained model while keeping the benefits. 

\subsection{On-The-Fly Audio Data Augmentation}
\label{ssec:dataaug}

We developed an on-the-fly data augmentation pipeline that gives each incoming audio example a probability of being altered. Alterations include varying pitch, volume, and speed, adding noise, randomly zeroing frames, bandpass filters, resampling, mixup~\cite{zhang2017mixup, xu2018mixupxception}, and negative data augmentation (NDA)~\cite{sinha2021negative}. 

Unlike previous applications of mixup where the proportion of mix of two examples is reflected in the labels, we consider two mixed items to have positive labels for all classes mixed together. For example, mixing an example of speech with an example of an engine at a ratio of .6/.4 will result in both classes having a label of 1 rather than reflecting mix ratio. 

We made NDA audio-specific rather than directly using image-based NDA. Flipped spectrograms, which are often label-preserving in images, are converted to negative examples. However, we create negative examples from: shuffled spectrograms (where the frequency bins or the time bins are shuffled randomly), jigsaw (2D areas are shuffled randomly), and cutout (randomly zeroing a portion of the spectrogram). The motivation of NDA is to encourage the model to learn global features rather than very local ones. 

\section{Models, Features, and Training Pipeline}
\label{sec:modeltraining}

For our experiments, we use the Xception model architecture as described in \cite{chollet2017xception}, which is a depthwise-separable CNN with residual connections. The full model can be seen reproduced from \cite{chollet2017xception} in Figure~\ref{fig:xception}. We also introduce an Xception-small model (Xception-s) that is identical to Xception but does not repeat the middle flow convolution layers. This change reduces the model size from 21 million parameters to 8 million. A comparison of the Xception models' parameter sizes and other top-performing models on ESC-50 is shown in Table~\ref{tab:modelsize}. 

The input features for all Xception and Xception-small models are a single-channel 80-bin log-mel filterbank. All audio was resampled to 16kHz. The two-dimensional shape of filterbanks provide some analogous properties to image recognition. The final output layer size is determined by the number of classes in each dataset. The final layer is 50 for ESC-50 and 527 for AudioSet. 

For training on the ESC-50 dataset, we use adam optimizer with an initial learning rate of 0.001 with a decay of 0.8 every epoch for 25 epochs. We use cross-entropy loss. The training process is repeated 5 times while rotating the training folds and evaluation fold as required by the ESC-50 leaderboard criterion. Data augmentation as described in Section \ref{ssec:dataaug} is applied but without mixup or NDA due to the single-label structure of the dataset. 

For pretraining on AudioSet, the same configuration of optimizer and learning rate is used. However, we use binary cross-entropy loss because AudioSet is a multi-label dataset where many examples include more than one positive labels. Data augmentation includes mixup and NDA. The AudioSet evaluation loss and mAP score is tracked, and training is stopped when the evaluation loss and mAP stagnates and the learning rate decay has passed below 1e-6. The epoch with the lowest evaluation loss is selected as the model to test. We also apply class weights to the criterion due to the strong imbalance in the AudioSet classes.

To convert an AudioSet model for fine-tuning on ESC-50, we remove the prediction layer from the AudioSet-trained model and apply a randomly-initialized linear layer of size 50. During fine-tuning, all parameters are updated rather than only updating the final layer(s). We have found this provides a higher accuracy on ESC-50 than freezing internal layers. 

\begin{table}[th]
  \centering
  \begin{tabular}{l c}
   \multicolumn{1}{c}{\textbf{Model}} & 
    \multicolumn{1}{c}{\textbf{N. Params (mil)}}\\
    \hline
     BigSSL-XXL~\cite{zhang2021bigssl} &  1000 \\
     \hline
     AST~\cite{gong2021ast} &  87  \\
     \hline
     PANNs~\cite{kong2020panns} &  81 \\
     \hline
     Xception (ours) &  21  \\
     Xception-s (ours) &  8 \\
     \hline
 \end{tabular}
  \caption{Comparison of model sizes and types.}
  \label{tab:modelsize}
  \end{table}

\section{Results}
\label{sec:majhead}
The full results of the Xception models with various configurations of knowledge transfer, pretraining with AudioSet, and data augmentation can be found in Table~\ref{tab:performance} along with other recent top-performing models. Visualizations of the average validation accuracy (averaged 5 folds per ESC-50 policy) for the first 25 epochs are shown in Figure~\ref{fig:res}. The average validation losses for the first 25 epochs are shown in Figure~\ref{fig:resloss}. 

\begin{table}[th]
  \centering
  \begin{tabular}{l c c c}
   \multicolumn{1}{c}{\textbf{Model}} & 
    \multicolumn{1}{c}{\textbf{Pretr.}} & 
    \multicolumn{1}{c}{\textbf{Aug.}} & 
    \multicolumn{1}{c}{\textbf{ESC-50 acc.}} \\
    \hline
     AST (SOTA) & IN + AS  & AS + ESC & 95.6  \\
     \hline
     PANNs & AS  & AS & 94.7  \\
     \hline
     BigSSL-XXL & UL & - & 90.9 \\
     \hline
     Xception-s 1 & -  & - & 76.3  \\
     Xception-s 2 & - & ESC & 77.4 \\ 
     Xception-s 3 & AS & - & 92.0 \\ 
     Xception-s 4 & AS & AS & 94.2 \\ 
     \hline
     Xception 1 & -        & -        & 72.5 \\
     Xception 2 & -        & ESC      & 72.5 \\
     Xception 3 & IN       & -        & 86.8  \\
     Xception 4 & IN       & ESC      & 86.1  \\
     Xception 5 & AS       & -        & 92.7 \\
     Xception 6 & AS       & AS        & 89.9 \\
    Xception 7 & IN + AS & - & 95.6 \\
     \textbf{Xception 8} & \textbf{IN + AS}  & \textbf{AS} & \textbf{95.8}  \\
 \end{tabular}
  \caption{Accuracy scores from the ESC-50 dataset compared with other top-scoring models. Comparing pretraining options: ImageNet weights (IN), AudioSet (AS); and augmentation options: AS, ESC-50 (ESC), and Unlabeled (UL). The Xception model pretrained with IN and AS with data augmentation during AS pretraining achieves SOTA results.}
  \label{tab:performance}
  \end{table}

\subsection{Xception}
\begin{figure}[htb]

\begin{minipage}[b]{1.0\linewidth}
  \centering
  \centerline{\includegraphics[width=7.5cm]{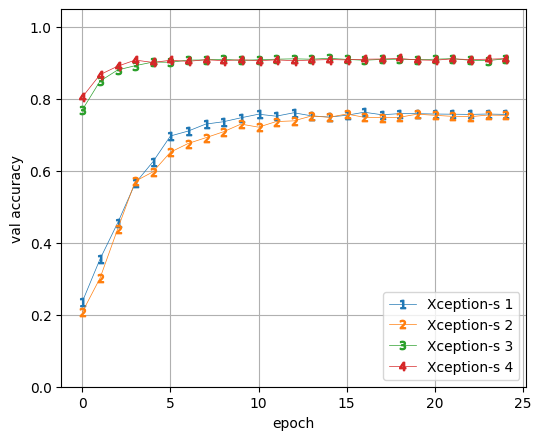}}
  \centerline{(a) Xception-small model accuracy}\medskip
\end{minipage}

\begin{minipage}[b]{1.0\linewidth}
  \centering
  \centerline{\includegraphics[width=7.5cm]{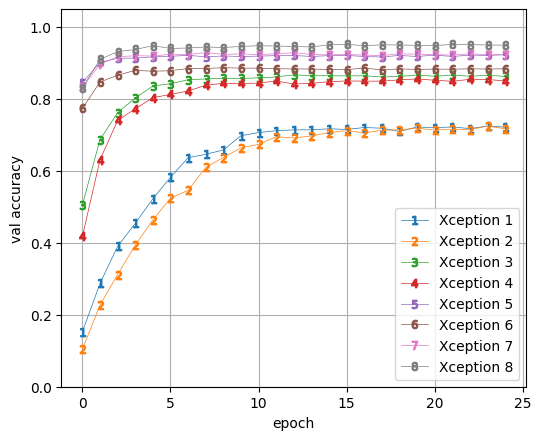}}
  \centerline{(b) Xception model accuracy}\medskip
\end{minipage}

\caption{Average validation accuracy over 5 folds of ESC-50 over 25 epochs for Xception and Xception-small models.}
\label{fig:res}
\end{figure}

\begin{figure}[htb]

\begin{minipage}[b]{1.0\linewidth}
  \centering
  \centerline{\includegraphics[width=7.5cm]{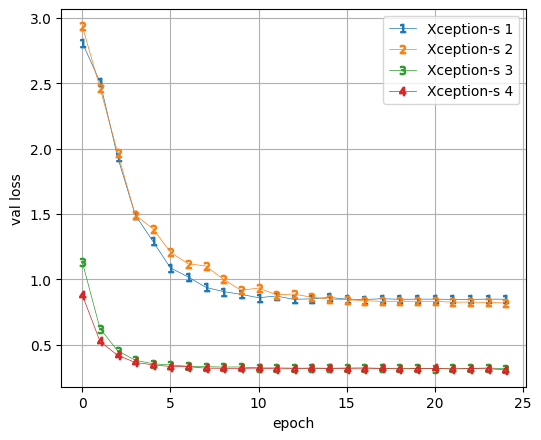}}
  \centerline{(a) Xception-small validation loss}\medskip
\end{minipage}

\begin{minipage}[b]{1.0\linewidth}
  \centering
  \centerline{\includegraphics[width=7.5cm]{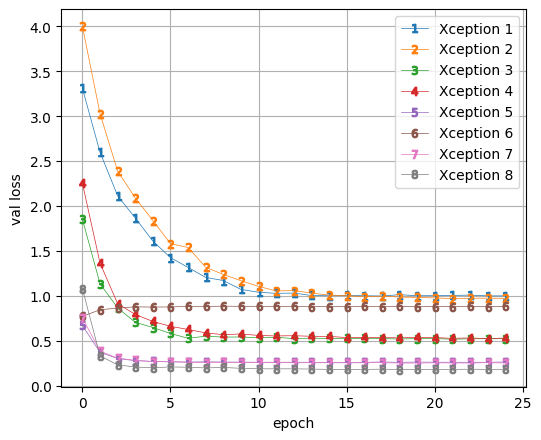}}
  \centerline{(b) Xception validation loss}\medskip
\end{minipage}

\caption{Average validation loss over 5 folds of ESC-50 over 25 epochs for Xception and Xception-small models.}
\label{fig:resloss}
\end{figure}

 Xception 8, which was initialized with ImageNet weights, pretrained with AudioSet, which was had on-the-fly augmentation, reaches SOTA. However, most of the models pretrained with AudioSet were close to SOTA such that small alterations to the ESC-50 fine-tuning pipeline could change the order of ranking. When the Xception model was pretrained with AudioSet, there were no performance gains from augmenting the ESC-50 data during fine-tuning. There is clear performance differences between models trained from scratch (Xception 1, 2), which scored the lowest, models initialized with ImageNet weights only (Xception 3, 4), which performed over 10 points higher and the remaining models pretrained with AudioSet, which achieve the highest scores. Initializing with ImageNet weights followed by training with AudioSet (Xception 7, 8) yields a slightly higher accuracy than only pretraining on AudioSet. Figure~\ref{fig:res} clearly shows this separation not only with accuracy scores but how quickly each model reaches optimal accuracy. These results are also reflected in the running validation losses in Figure~\ref{fig:resloss}. 

\subsection{Xception-small}
Similar to the Xception models, the Xception-small models that were pretrained with AudioSet scored much higher than those trained from scratch. The best Xception-small model scores near the other top-scoring models despite having a tenth or fewer parameters. While the full Xception model did not show any impact with data augmentation, the Xception-small models showed modest improvement in performance by including data augmentation. When comparing training on ESC-50 from scratch, Xception-small performs better than the full Xception model, likely because of it has fewer parameters and is less likely to over-fit.

\section{Conclusion and Further Work}
We have shown that Xception models can reach SOTA on ESC-50 when they are pretrained with AudioSet and are much smaller than other top-performing models. In the absence of a larger in-domain dataset, applying knowledge transfer from an outside domain such as image recognition gives a better result than training on a small dataset directly. Data augmentation only benefits the Xception-small model, and only by a small amount. 

Future work will include evaluating on other datasets, especially multi-label datasets. We would also like to analyze the individual components of the data augmentation pipeline to find which, if any, are most beneficial. Furthermore, we would like to attempt knowledge transfer from other domains than image recognition. 

Further work should be done in reducing the size of models, such as converting Xception to Xception-small, while retaining the benefit of pretrained ImageNet weights. Deleting the layers directly removed any benefit of the ImageNet models. 

Additional study of model performance relative to parameter size and runtime on low-compute devices would be beneficial, especially with the increase of running AED models on low-compute devices.

\vfill\pagebreak



\bibliographystyle{IEEEbib}
\bibliography{strings,refs}

\end{document}